\documentclass[journal]{IEEEtran}
\usepackage{cite}
\usepackage{amsmath,amssymb,amsfonts}
\usepackage{algorithmic}
\usepackage{graphicx}
\usepackage{textcomp}
\usepackage{xcolor}
\usepackage{booktabs} 
\newcommand{\tabincell}[2]{\begin{tabular}{@{}#1@{}}#2\end{tabular}} 
\def\BibTeX{{\rm B\kern-.05em{\sc i\kern-.025em b}\kern-.08em
    T\kern-.1667em\lower.7ex\hbox{E}\kern-.125emX}}

\ifCLASSINFOpdf
\else
\fi
\begin{document}
%
\title{Diving Body Shape Coefficient Setting Based on Moment of Inertia Analysis}
%
%
%

\author{Yan-Xin~Sun,
        and~Zhao-Hui~Sun,~\IEEEmembership{Member, IEEE}
\thanks{Y.X. Sun is with the School
of Mechanical Engineering, Shanghai Jiao Tong University, 200240 Shanghai, China.
 (E-mail: sjtu$\_$syx@163.com).}
\thanks{Z.H. Sun is with the School
of Mechanical Engineering, Shanghai Jiao Tong University, 200240 Shanghai, China
and also with SJTU Innovation Center of Producer Service Development, Shanghai Research Center for Industrial Informatics, Shanghai Key Lab of Advanced Manufacturing Environment, Institute of Intelligent Manufacturing. (E-mail: zh.sun@sjtu.edu.cn).}
\thanks{This work was sponsored by National Natural Science Foundation
of China (Grant No. 71632008).}
}

\maketitle

\begin{abstract}
In the diving competition rules, FINA specifies the code of different diving movements and its difficulty coefficient. The rule simply relies on the complexity of the action to determine the difficulty. In the formulation of the diving difficulty coefficient, the athlete's body shape has not been fully considered, so it is difficult to fully guarantee the fairness of the diving competition. Based on the above problems, this paper analyzes the rules of the FINA's 10-meter platform diving difficulty coefficient, establishes the multi-rigid-body model of the human body, obtains the relationship between the moment of inertia and the completion time of the athletes to complete each diving action and the athlete's body shape, and determines the index to measure the athlete's body shape. The Lagrange Interpolation Polynomial is used to establish the functional relationship between the body shape correction coefficient and the body shape correction index, and the body shape correction coefficient corresponding to different body type athletes is determined accordingly. Finally, a new 10-meter platform diving difficulty coefficient scheme was developed. 
\end{abstract}


\section{Introduction} 
In the diving competition rules, FINA specifies the codes of different diving actions and their difficulty coefficients, which are related to the diving player's take-off mode and air movement. When judging the score, according to the performance of the athletes' performance and the water intake effect, they give the action score from 10 to 0, and then calculate the completion score of the athlete according to a certain formula. And the product of the completion score and the difficulty coefficient of the action is the final score of the athlete's action. The typical diving action consists of a series of somersault and twisting movements. The athletes generate sufficient angular momentum when they take off, complete each diving action by changing postures in the air, and finally enter the water vertically. Frohlich$^{[1]}$ first gave the correct twisting somersault physics model and pointed out that the change of athlete's posture has an important influence on the change of rotational angular velocity. Later, Yeadon analyzed the process of athletes tumbling in the air in [2-5], and more comprehensively analyzed the mechanical model of the twisting somersault in [6-9]. Sudarsh Bharadwaj$^{[10]}$ et al. proposed a simple twisting somersault model consisting of a rigid body and a rotor. The model derived a clear formula and explained how to formulate a set of diving action based on the formula. Holger R. Dullin$^{[11]}$ et al. simplified the Euler equations for non-rigid dynamics, gave a dynamic analysis of the twisting somersault, and derived a precise formula for the twisting somersault to calculate the angular momentum, the time spent, and the energy consumed in each diving stage of the diver, and the total time required for the diver to complete a set of diving exercises is calculated accordingly.  William Tong$^{[12]}$ et al. proposed a coupled rigid body model and applied the Euler equation to the non-rigid-body field. The detailed analysis of the difficult diving action 513XD was carried out to demonstrate the completeness of the action and propose the completion plan of this action. The modeling analysis of this paper finds that the body shape of the diving athlete will directly affect the completion time of the diving action, which will affect the difficulty of diving. The current diving difficulty coefficient rule does not consider the impact of the athlete's body shape on the diving difficulty, that is, athletes of different body shape complete the same diving action, the corresponding difficulty coefficient is the same. This paper believes that there is a certain irrationality in this rule. In order to ensure the fairness of diving, the body shape coefficient should be set to correct and eliminate the advantages of slim and small athletes.

The structure of this paper is as follows: In Section 2, this paper analyzes the current diving movement difficulty coefficient rules, obtains the regular pattern of the current rules, and proposes model assumptions to lay the foundation for the process of modeling later. In Section 3, this paper establishes the multi-rigid-body model of the human body, obtains the completion time of full somersault in pike position, full somersault in pike position and full twisting accordingly, and analyzes the relationship between the completion time of the diving action and the body shape of the athletes. In Section 4, this paper formulates the body shape correction coefficient rule according to the relationship between the completion time of the diving action and the athlete's body shape, and formulates the basic action difficulty coefficient based on the completion time of a set of diving action, and introduces the expert evaluation coefficient to further correct the new difficulty coefficient rule. According to the above coefficients, a new diving action difficulty coefficient rule is formulated and compared with the old rules. In Section 5, this paper simulates the newly established diving difficulty coefficient rule, and shows the use of the new diving action difficulty coefficient rule by example.

\section{Analysis of Current Rules and Model Assumptions} 

\subsection{Analysis of the Current Rule} 

The FINA’s method for determining the difficulty coefficient of diving action is given in APPENDIX 3, 4 of FINA Diving Rules. This paper summarizes the following rules:

1) For the same set of diving actions, the difficulty coefficient of somersault in pike position (\emph{B}) is higher than the difficulty coefficient of somersault in tuck position (\emph{C}).

2) The more the number of somersault and twisting, the higher the difficulty coefficient of the corresponding action is.

3) When the number somersault and twisting are the same, if the athlete's take-off mode and the direction of somersault are different, the difficulty of the action is different.
It can be seen from the above rules that in the 10-meter platform diving, the more complicated the diving action is completed in the limited air time, the more number of somersault and twisting there are, the higher the difficulty coefficient of a set of diving action is.

\subsection{Research Hypothesis} 

\begin{itemize}
\item Symbol description
\end{itemize} 

The symbols used in this article and their meanings are shown in Table \uppercase\expandafter{\romannumeral1}. 

\begin{table*}[htbp]
\centering
\caption{Symbols used in this article and their meaning}
\label{T2-1}
\begin{tabular}{cc}
\toprule  
Symbols& Description\\
\midrule  
\emph{M} & Mass of the athlete\\
\emph{h} & Height of the athlete\\
\emph{m} & \emph{1/100M}\\
\emph{a} & \emph{1/16h}\\
\emph{J} & Initial angular momentum\\
\emph{$\omega$} & Angular velocity\\
\emph{I$_{i}$} & Moment of inertia of the \emph{i}-th part\\
\emph{I$_{B}$} & Total moment of inertia in pike position\\
\emph{I$_{C}$} & Total moment of inertia in tuck position\\
\emph{I$_{T}$} & Total moment of inertia of twisting\\
\emph{t$_{B}$} & Completion time of full somersault in pike position\\
\emph{t$_{C}$} & Completion time of full somersault in tuck position\\
\emph{t$_{T}$} & Completion time of full twisting\\
\emph{Mh$^{2}$\,$_{min}$} & Minimum of body shape correction coefficient\\
\emph{Mh$^{2}$\,$_{avr}$} & Average of body shape correction coefficient\\
\emph{Mh$^{2}$\,$_{max}$} & Maximum of body shape correction coefficient\\
\emph{DD} & New action difficulty coefficient\\
\emph{DD'} & Action difficulty coefficient after expert evaluation coefficient correction\\
\emph{DD''} & Basic action difficulty coefficient\\
\emph{BC} & Body shape correction coefficient\\
\emph{EA} & Expert evaluation coefficient\\
\bottomrule 
\end{tabular}
\end{table*}

Note: Other symbols are described in the text.

\begin{itemize}
\item Model hypothesis
\end{itemize} 

In order to simplify the model and solution process, make the model simple and reasonable, this paper makes the following assumptions:

1) The human body is a multi-rigid-body model, various parts of the body are regarded as rigid bodies, and each rigid body can be relatively moved to complete the diving action.

2) Different athletes, after adequate training, could produce the same initial angular momentum in different take-off positions and somersault modes.

3) The change of body posture during the completion of the diving action is instantaneous$^{[13]}$, so the athlete's angular momentum is conserved during this process.

4) The movement of the athlete in the height direction is a free fall motion with an initial velocity of 0, and the gravitational acceleration g = 9.8 m/s$^{2}$.

5) Remove the 0.1s take-off time and 0.1s entry time, the rest of the time are used to complete the diving action.

6) The athlete does not tilt after taking off.

7) During the process of somersault and twisting, the athlete's axis of rotation passes through the center of mass.

\section{Modeling Process} 

\subsection{Multi-rigid-body Model of Human Body} 

In this paper, the human body is divided into six parts: the trunk, the head and the limbs to establish the model$^{[14,15]}$, and each part is simplified into a rectangular shape, as shown in Fig. 1. 

\begin{figure}[ht]

\centering
\includegraphics[scale=0.6]{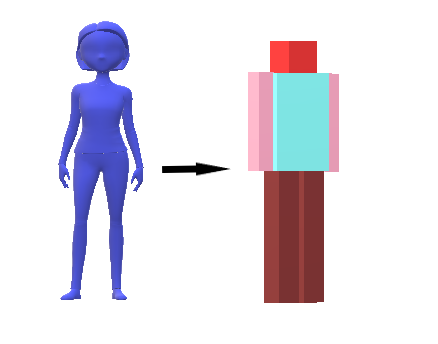}
\caption{Schematic diagram of human body model}
\label{F3-1}
\end{figure}

\begin{figure}[ht]

\centering
\includegraphics[scale=0.7]{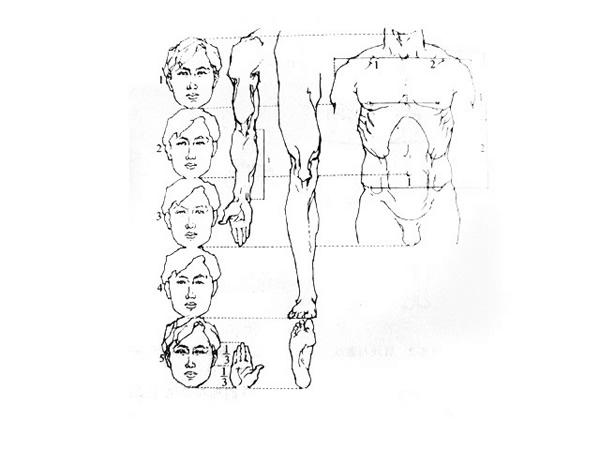}
\caption{Human body size scale}
\label{F3-2}
\end{figure}

Assume that the total height of the athlete is \emph{h=16a}, and the proportion of the human body is analyzed. As shown in Fig. 2, the length of the trunk is about 3/8 of the total height, that is, \emph{6a}, and the length of the head is about 1/8 of the total height, that is, \emph{2a}, the length of the arm is the same as the length of the trunk, that is, \emph{6a}, and the length of the leg is about 1/2 of the total height, that is \emph{8a}. The mass ratio of various parts of the human body$^{[16,17]}$ is: the head accounts for 8\% of the total mass of the human body, the single arm accounts for 5\%, the trunk accounts for about 50\%, and the single leg accounts for about 16\%. Assume that the athlete's mass is \emph{M=100m}, the mass of the athlete's head, single arm, trunk and single leg are \emph{8m}, \emph{5m}, \emph{50m} and \emph{16m} respectively. It is assumed that the density of each part of the body of the athlete is evenly distributed. According to the formula (3-1), the volume ratio of each part of the human body is equal to the mass ratio. 

\[
M=\rho V\eqno(3-1) 
\]

In this paper, the human body is divided into six cuboids. The length of each part is shown in Fig. 2. It is assumed that the head of the human body is a square body, the width of the limbs is equal to the thickness, and the width of the trunk is twice the thickness. According to the above assumption, the quality and size of each part of the human body is obtained, as shown in Table \uppercase\expandafter{\romannumeral2}. 

\begin{table}[!htbp]
\centering
\caption{Quality and size setting table of the human body}
\label{T3-1}
\begin{tabular}{ccccc}
\toprule
Parts of body & Length & Width & Thickness & Mass \\
\midrule
Head & \emph{2a} & \emph{2a} & \emph{2a} & \emph{8m} \\
Arm & \emph{6a} & \emph{a} & \emph{a} & \emph{6m} \\
Trunk & \emph{6a} & \emph{4a} & \emph{2a} & \emph{48m} \\
Leg & \emph{8a} & \emph{1.4a} & \emph{1.4a} & \emph{16m} \\
\bottomrule
\end{tabular}
\end{table}

This model is mainly used to analyze the relationship between the time when athletes complete each diving action and the athlete's body type (height, weight). Based on the multi-rigid-body model of the human body, this paper will further establish the corresponding action model based on the decomposition of a set of diving actions. According to Hypothesis 4), the angular momentum of the athlete is conserved during the completion of the diving action, and the main actions of the diving can be divided into the somersault and twisting actions, which can be regarded as the rotation motion of rigid body. According to the calculation method of the moment of inertia$^{[18,19]}$: set the mass of the rigid body to \emph{m$_{i}$}, the moment of inertia around the rotation axis is \emph{I$_{i}$'} , and move the axis parallel by a distance \emph{d$_{i}$}, then the moment of inertia around the new axis \emph{I$_{i}$} is: 

\[
I_i=I_i'+I_i''\eqno(3-2) 
\]

For the cuboid, when its rotary axis is its central axis, its moment of inertia is \emph{I$_{i}$'}, \emph{l$_{1}$}, \emph{l$_{2}$} are the lengths of the two sides of the rectangle perpendicular to the axis of rotation. 

\[
I_i'=\frac{1}{12}m_i(l_1^2+l_2^2)\eqno(3-3) 
\]
\[
I_i''=m_id_i^2\eqno(3-4)
\]


The total moment of inertia of the model is: 

\[
I=\sum_{i=1}^nI_i=\sum_{i=1}^n(I_i'+I_i'')=\sum_{i=1}^nI_i'+\sum_{i=1}^nI_i''\eqno(3-5) 
\]

\subsection{Modeling of Diving Action} 

For somersault in pike position, somersault in tuck position and twisting action, combined with the multi-rigid-body model of the human body, follow the steps below to analyze and solve.

Step 1: According to the multi-rigid-body model of the human body, the human body is divided into six rigid bodies.

Step 2: Further correct the pose of the model according to the actual pose of somersault in pike position, somersault in tuck position and twisting action, and establish an appropriate coordinate system. The diving action model established in this paper is shown in Fig. 3.

Step 3: Using Newton's classical mechanics to derive the relationship between the completion time of the diving action and the athlete's height and weight. The derived conclusions will be used to develop the body shape correction factor. 

\begin{figure*}[ht]

\centering
\includegraphics[scale=0.8]{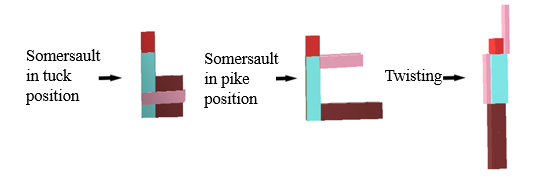}
\caption{Model of the diving action}
\label{F3-3}
\end{figure*}

The first action is the somersault in pike position. Considering that the somersault action is rotating around the Y-axis, the coordinate system shown in Fig. 4 is established, and the model centroid coordinates are (\emph{X$_{C}$, 0, Z$_{C}$}). 

\begin{figure}[ht]

\centering
\includegraphics[scale=0.7]{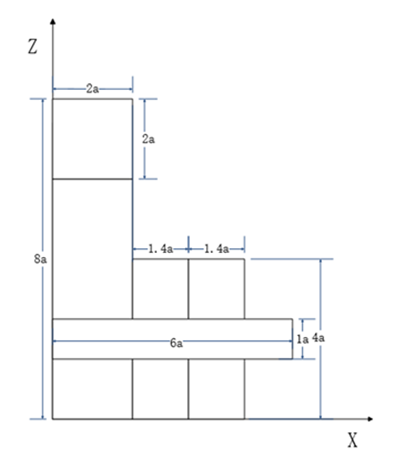}
\caption{Model of the somersault in tuck position}
\label{F3-4}
\end{figure}

According to the multi-rigid body model of the human body, it is assumed that the total mass of the athlete is \emph{M=100m}, and the total height \emph{h=16a}. 

\[
X_C=\sum_{i=1}^6\frac{x_im_i}{M}\eqno(3-6)
\]
\[
Z_C=\sum_{i=1}^6\frac{z_im_i}{M}\eqno(3-7)
\]

Through the centroid coordinate formula (3-6), (3-7), the centroid coordinate \emph{X$_{C}$=2.008a} and \emph{Z$_{C}$=2.88a} are calculated. Therefore, the centroid coordinates are (\emph{2.008a, 0, 2.88a}).

According to the calculation formula of the moment of inertia (3-8), (3-9): 

\[
I_C'=\sum_{i=1}^6\frac{1}{12}m_i(l_{1i}^2+l_{2i}^2)\eqno(3-8) 
\]
\[
I_C''=\sum_{i=1}^6m_i[(x_i-X_C)^2+(z_i-Z_C)^2]\eqno(3-9) 
\]
Calculated: 

$I_C'=250.2267ma^2$

$I_C'' =316.9536ma^2$

$I_C=567.1803ma^2=0.0222Mh^2$

Therefore, the time taken by the athlete to complete a full somersault in tuck position is: 

\[
t_C=\frac{2\pi I_C}{J}=\frac{2\pi\times0.0222Mh^2}{J}\eqno(3-10) 
\]

In the same way, the analysis of the somersault in pike position is assumed to be the same as the above analysis. Establish the coordinate system as shown in Fig. 5, and set the model centroid coordinates to (\emph{X$_{B}$, 0, Z$_{B}$}): 

\[
X_B=\sum_{i=1}^6\frac{x_im_i}{M}\eqno(3-11) 
\]
\[
Z_B=\sum_{i=1}^6\frac{z_im_i}{M}\eqno(3-12)
\]

According to the centroid coordinate formula (3-11), (3-12), the centroid coordinates of the somersault in pike position are calculated as (\emph{2.884a, 0, 3.08a}).

\begin{figure}[ht]

\centering
\includegraphics[scale=0.8]{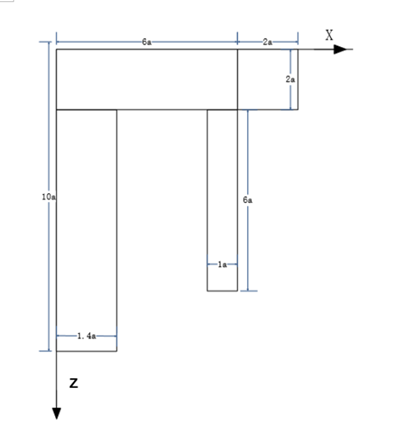}
\caption{Model of the somersault in pike position}
\label{F3-5}
\end{figure}

Similarly, the formula for calculating the moment of inertia is: 

\[
I_B'=\sum_{i=1}^6\frac{1}{12}m_i(l_{1i}^2+l_{2i}^2)\eqno(3-13) 
\]
\[
I_B''=\sum_{i=1}^6m_i[(x_i-X_B)^2+(z_i-Z_B)^2]\eqno(3-14)
\]

Calculated: 

$I_B'=378.2267ma^2$

$I_B''=930.2944ma^2$

$I_B=1308.5211ma^2=0.0511Mh^2$

Therefore, the time taken for an athlete to complete a full somersault in pike position is: 

\[
t_B=\frac{2\pi I_B}{J}=\frac{2\pi\times0.0511Mh^2}{J}\eqno(3-15) 
\]

Finally, the twisting action is analyzed. The twisting action is the rotation around the Z axis, and the component on the Z axis can be ignored (the Z axis is still marked in Fig. 6 in order to show the various parts of the human body, which is not counted in the actual calculation). Establish the coordinate system shown in Fig. 6, and set the model centroid coordinate to (\emph{0, Y$_{T}$, 0}). 

\[
Y_T=\sum_{i=1}^6\frac{y_im_i}{M}\eqno(3-16)
\]

According to the centroid coordinate formula (3-16), the centroid coordinate of the twisting action is calculated as \emph{Y$_{T}$ =0}, so the centroid coordinate is: (\emph{0,0,0}). 

\begin{figure}[ht]

\centering
\includegraphics[scale=0.6]{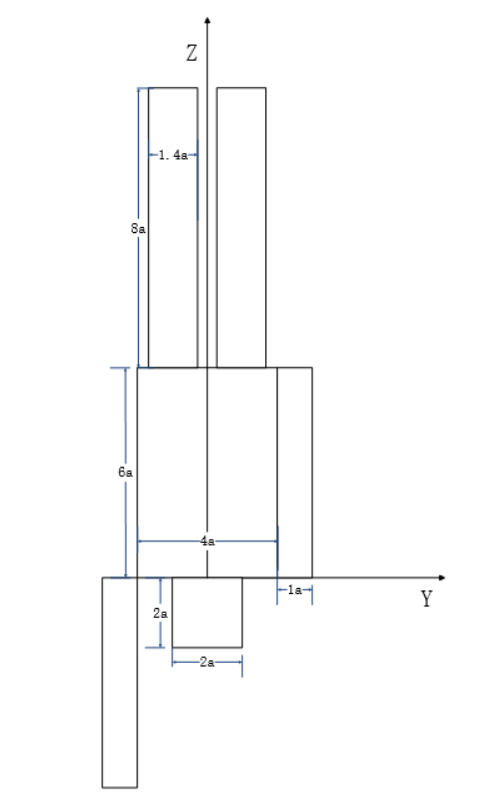}
\caption{Model of the twisting action}
\label{F3-6}
\end{figure}

Calculate the moment of inertia by equations (3-17), (3-18): 

\[
I_T'=\sum_{i=1}^6\frac{1}{12}m_i(l_{1i}^2+l_{2i}^2)\eqno(3-17) 
\]
\[
I_T''=\sum_{i=1}^6m_i(y_i-Y_T)^2\eqno(3-18) 
\]

Calculated: 

$I_T'=98ma^2$

$I_T''=90.68ma^2$

$I_T=188.68ma^2=0.0074Mh^2$

Therefore, the time taken for an athlete to complete a full twisting is: 

\[
t_T=\frac{2\pi I_T}{J}=\frac{2\pi\times0.0074Mh^2}{J}\eqno(3-19) 
\]

According to statistics, the time taken by Chinese female diver Ren Qian to complete the 309B action(4.5 somersaults in pike position) is 1.2s, and the follow-up calculation is completed according to Ren's body parameters (Note: Ren's height and weight are closer to the average of the diver): 

\[
4.5\times t_B=4.5\times\frac{2\pi\times0.0511Mh^2}{J}=1.2\eqno(3-20) 
\]

By formula (3-20), the reasonable value of the initial moment of inertia \emph{J} is calculated as: 

\[
J=142.19kg\cdot m^2\cdot s^{-1}\eqno(3-21) 
\]

Substituting the value of \emph{J} into the formulas (3-10), (3-15), and (3-19), the relationship between the time when the athlete completes each diving action and the height \emph{h} and the weight \emph{M} of the athlete is obtained as shown in the formula (3-22). 

\[ 
t_B=2.254\times 10^{-3}Mh^2
\]

\[
  t_C=9.810\times 10^{-4}Mh^2\eqno(3-22)
\]

\[ 
t_T=3.270\times 10^{-4}Mh^2
\]


The time ratio of each diving action is: 

\[
t_B:t_C:t_T\approx 7:3:1 
\]

The relationship between the completion time of each action and the weight and height is shown in Fig. 7 (the red, blue and green lines respectively represent the body weight \emph{M=40, 50, 60kg}), Fig. 8 (Red, blue, and green lines respectively represent weight \emph{h=1.5, 1.6, 1.7m}). 

\begin{figure}[ht]

\centering
\includegraphics[scale=0.65]{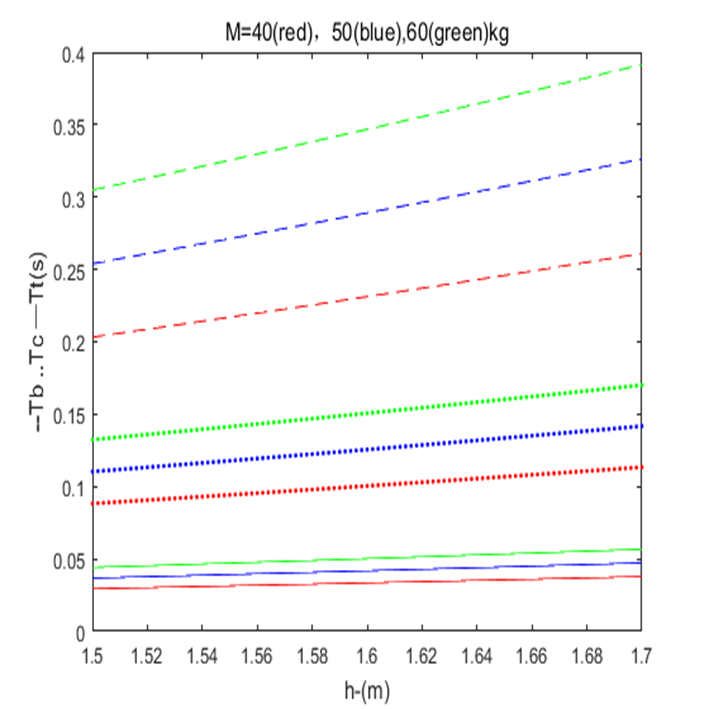}
\caption{Relationship between completion time of each action and height}
\label{F3-7}
\end{figure}

\begin{figure}[ht]

\centering
\includegraphics[scale=0.4]{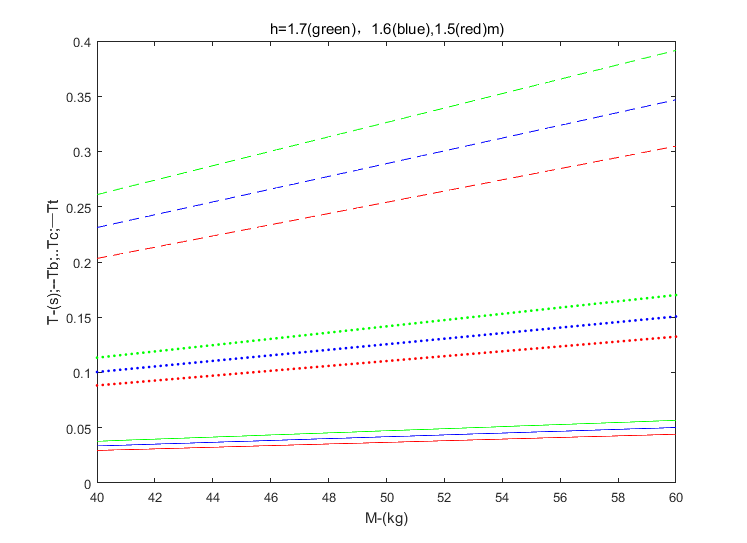}
\caption{Relationship between completion time of each action and weight}
\label{F3-8}
\end{figure}

It can be seen from Fig. 7 and Fig. 8 that the completion time of each action is linearly positively correlated with the body weight, and is positively correlated with the height, which is the same as conclusion obtained by the model.

Then, taking action 309B as an example, the relationship between the time required to complete the entire 309B action and the height and weight is obtained, as shown in Fig. 9 (the red, blue and green lines respectively represent the body weight \emph{M=40, 50, 60kg}), Fig. 10 (red, blue, and green lines respectively represent weight \emph{h=1.5, 1.6, 1.7m}). 

\begin{figure}[ht]

\centering
\includegraphics[scale=0.71]{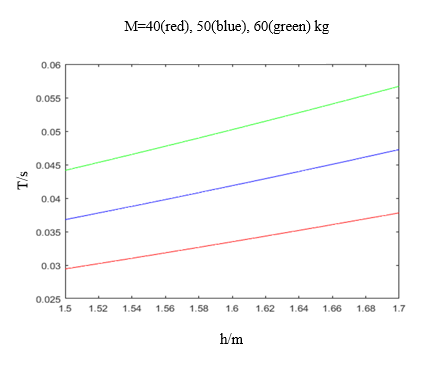}
\caption{Relationship between 309B action completion time and height}
\label{F3-9}
\end{figure}

\begin{figure}[ht]

\centering
\includegraphics[scale=0.8]{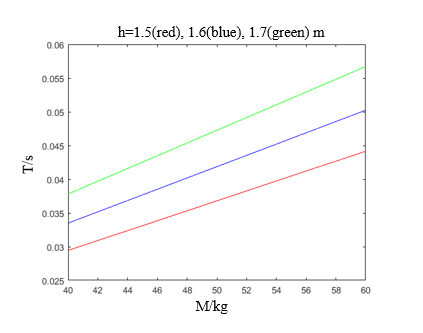}
\caption{Relationship between 309B action completion time and weight}
\label{F3-10}
\end{figure}

It can be seen from Fig. 9 and Fig. 10 that the 309B action completion time is linearly positively correlated with body weight, and is positively correlated with height, which is the same as conclusion obtained by the model.

Since the formula (3-22) reflects that the completion time of the action is proportional to \emph{Mh$^{2}$}, \emph{Mh$^{2}$} is used as a reference index of the body type correction parameter \emph{BC}, so the relationship between the completion time of each action and \emph{Mh$^{2}$} is also analyzed, as shown in Fig. 11. 

\begin{figure}[ht]

\centering
\includegraphics[scale=0.74]{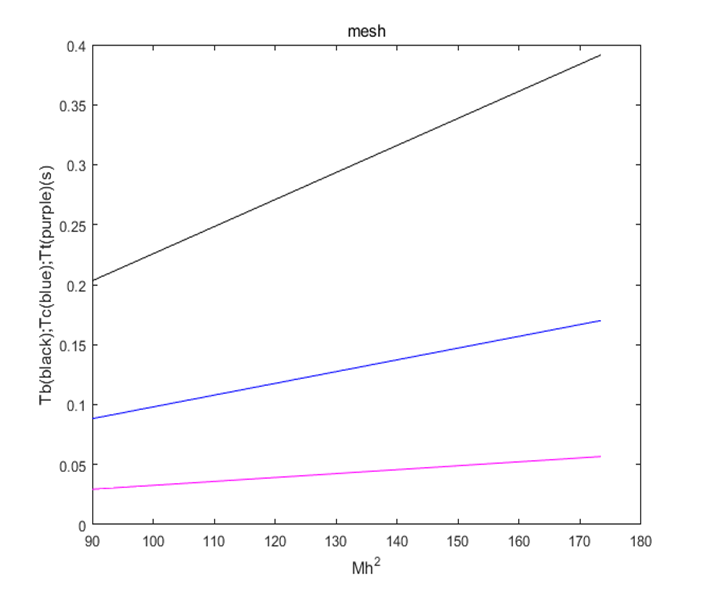}
\caption{Relationship between the completion time of each action and body shape correction index}
\label{F3-11}
\end{figure}

It can be seen from Fig. 11 that the completion time of each action is linearly positively correlated with \emph{Mh$^{2}$}, which is the same as the conclusion obtained by the model.

\section{Formulation of New Diving Difficulty Coefficient Rules} 

\subsection{Body Shape Correction Coefficient} 

From the relationship between the athlete's completion time of each diving action and the athlete's height \emph{h} and weight \emph{M}, that is, formula (3-22), the athlete's body shape has a direct impact on the completion time of the action. Therefore, the athlete's body shape and the action difficulty coefficient are directly related. The body shape correction coefficient should be set to correct the current diving action difficulty coefficient rule. The setting of the athlete's body shape correction coefficient should be considered from two aspects: First, the completion time of the diving action is proportional to \emph{Mh$^{2}$}.Therefore, \emph{Mh$^{2}$} is selected as the measurement index of the athlete's body shape, that is, the body shape correction index. The typical diving athlete's body shape correction index \emph{Mh$^{2}$} is obtained, which is used as a benchmark to ensure that the body shape correction coefficient has good applicability. Secondly, Lagrange Interpolation Polynomial is combined with the benchmark data to determine the body shape correction coefficient of athletes of different body size. Avoiding direct linear fit makes the body shape correction coefficient setting unreasonable.

According to formula (3-22), the time taken by the athlete to complete the diving action is directly proportional to the body shape correction index \emph{Mh$^{2}$}. The smaller the \emph{Mh$^{2}$} is, the shorter the time for completing a certain action is, and the easier it is to get a high action score. Only considering the body factor, assuming that the same diving action is completed, the action score of the athlete, whose body shape correction index \emph{Mh$^{2}$} is the maximum, is 6, the action score of the athlete, whose \emph{Mh$^{2}$} is average, is 8, and the action score of the athlete, whose \emph{Mh$^{2}$} is minimum, is 10. The introduced body shape correction coefficient \emph{BC} should give everyone the same score to eliminate the influence of the body factor. The initial setting scheme of the specific body shape correction coefficient is shown in Table \uppercase\expandafter{\romannumeral3}. 

\begin{table}[!htbp]
\centering
\caption{Initial setting table of body type correction coefficient}
\label{T4-1}
\begin{tabular}{cccc}
\toprule
 & \emph{Mh$^{2}$\,$_{min}$} & \emph{Mh$^{2}$\,$_{avr}$} & \emph{Mh$^{2}$\,$_{max}$}  \\
\midrule
Original action score & 10 & 8 & 6 \\
\emph{BC}(body shape correction index) & 0.8 & 1 & 1.33 \\
Final score & 8 & 8 & 8 \\
\bottomrule
\end{tabular}
\end{table}

According to the typical diving athletes' body size data, the minimum, average and maximum values of the female diving athlete's body shape correction index are \emph{$98.596kg\cdot m^{2}$}, \emph{$121.4991kg\cdot m^{2}$} and \emph{$140.8157kg\cdot m^{2}$}. The minimum, average and maximum values of the male diving athlete's body shape correction index are \emph{$107.52kg\cdot m^{2}$}, \emph{$164.5124kg\cdot m^{2}$}, and \emph{$207.6111kg\cdot m^{2}$}. It could be seen that the distribution of body shape correction indicators for male and female diving athletes is very different. Therefore, this paper formulates body shape correction coefficient for male and female diving athletes respectively.

In this paper, Lagrange Interpolation Polynomial is used to determine the relationship between the body shape correction coefficient \emph{BC} of the diving athlete and the body shape correction index \emph{Mh$^{2}$}. The minimum, average and maximum values of the female diver's body shape correction index are \emph{$98.596kg\cdot m^{2}$}, \emph{$121.4991kg\cdot m^{2}$} and \emph{$140.8157kg\cdot m^{2}$}. So the coordinates of the three interpolation points are (\emph{98.596, 0.8}), (\emph{121.4991, 1}), (\emph{140.8157, 1.33}), which is substituted into the Lagrange interpolation formula to calculate the relationship between the body shape correction coefficient \emph{BC} of the female diving athlete and the body shape correction index \emph{Mh$^{2}$}(\emph{x=Mh$^{2}$}): 

\[
L_1(x)=2\times 10^{-4}x^2-0.035x+2.3\eqno(4-1) 
\]
\[
BC=2\times 10^{-4}M^2h^4-0.035Mh^2+2.3\eqno(4-2)
\]

The segmented integrals of the body shape correction index are averaged to obtain the body shape correction coefficient of the female diving athletes, as shown in Table \uppercase\expandafter{\romannumeral4}. 

\begin{table}[!htbp]
\centering
\caption{Body shape correction coefficient table of female athlete}
\label{T4-2}
\begin{tabular}{ccccc}
\toprule
 & \emph{Mh$^{2}/kg\cdot m^{2}$} & & \emph{BC} &  \\
\midrule
 & 90-100 &  & 0.79 &  \\
 & 100-110 & & 0.84 &  \\
 & 110-120 & & 0.93 &  \\
 & 120-130 & & 1.05 &  \\
 & 130-140 & & 1.22 &  \\
 & 140-150 & & 1.43 &  \\
\bottomrule
\end{tabular}
\end{table}

The male athlete's body shape data is treated in the same way, and the male diving athlete's body shape correction coefficient scheme can be obtained, as shown in Table \uppercase\expandafter{\romannumeral5}. 

\begin{table}[!htbp]
\centering
\caption{Body shape correction coefficient table of male athlete}
\label{T4-3}
\begin{tabular}{ccccc}
\toprule
 & \emph{Mh$^{2}/kg\cdot m^{2}$} & & \emph{BC} &  \\
\midrule
 & 100-110 & & 0.83 &  \\
 & 110-120 & & 0.85 &  \\
 & 120-130 & & 0.87 &  \\
 & 130-140 & & 0.89 &  \\
 & 140-150 & & 0.93 &  \\
 & 150-160 & & 0.98 &  \\
 & 160-170 & & 1.03 &  \\
 & 170-180 & & 1.09 &  \\
 & 180-190 & & 1.16 &  \\
 & 190-200 & & 1.24 &  \\
 & 200-210 & & 1.32 &  \\
\bottomrule
\end{tabular}
\end{table}

The relationship between body shape correction coefficient and body shape correction index is plotted, as shown in Fig. 12. 

\begin{figure}[ht]

\centering
\includegraphics[scale=0.75]{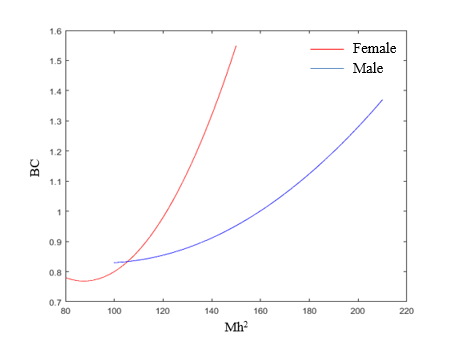}
\caption{Relationship between body shape correction coefficient and body shape correction index}
\label{F4-1}
\end{figure}

It could be seen that the larger the athlete's body shape correction index \emph{Mh$^{2}$} is, the larger the corresponding body shape correction coefficient is, and the body shape correction coefficient achieves the purpose of eliminating the advantage of the slim type athlete.

\subsection{Basic Action Difficulty Coefficient} 

According to the conclusions obtained by the formula (3-22), it could be seen that the time ratio of the same athlete completing full somersault in pike position, full somersault in tuck position, and full twisting is about \emph{7:3:1}. Assume that the time to complete full twisting is 1, the time to complete full somersault in tuck position is 3, and the time to complete full somersault in pike position is 7, then the time taken to complete the action code 105B (2.5 somersaults in pike position) is 17.5. According to the above rules, calculate the action completion time of all action codes. Search the 10-meter platform difficulty coefficient table, and find that the action code with the lowest difficulty coefficient is 105C, the difficulty coefficient is 2.1, and the completion time is 7.5. The action code with the highest difficulty coefficient is 309B, the difficulty coefficient is 4.8, and the completion time is 31.5. Taking these two actions as interpolation points, the linear equation of the basic difficulty coefficient with respect to the completion time of the action is obtained by the linear interpolation, as shown in formula (4-3): 

\[
DD''=\frac{9}{80}+\frac{201}{160}\eqno(4-3) 
\]

The different action completion times are substituted into the above formula, and the basic difficulty coefficient \emph{DD''} corresponding to each action code is obtained, as shown in Table \uppercase\expandafter{\romannumeral6}. 

\begin{table}[!htbp]
\centering
\caption{10-meter platform basic action difficulty coefficient}
\label{T4-4}
\begin{tabular}{c|cc|cc}
\toprule
 &\multicolumn{2}{c}{B}&\multicolumn{2}{|c}{C} \\
\midrule
Action code & Completion time & \emph{DD''} & Completion time & \emph{DD''} \\
\midrule
105 & 17.5 & 3.23 & 7.5 & 2.1 \\
107 & 24.5 & 4.01 & 10.5 & 2.44 \\
109 & 31.5 & 4.8 & 13.5 & 2.78 \\
1011 & 38.5 & 5.59 & 16.5 & 3.11 \\
205 & 17.5 & 3.23 & 7.5 & 2.1 \\
207 & 24.5 & 4.01 & 10.5 & 2.44 \\
209 & 31.5 & 4.8 & 13.5 & 2.78 \\
305 & 17.5 & 3.23 & 7.5 & 2.1 \\
307 & 24.5 & 4.01 & 10.5 & 2.44 \\
309 & 31.5 & 4.8 & 13.5 & 2.78 \\
405 & 17.5 & 3.23 & 7.5 & 2.1 \\
407 & 24.5 & 4.01 & 10.5 & 2.44 \\
409 & 31.5 & 4.8 & 13.5 & 2.78 \\
5154 & 19.5 & 3.45 & 9.5 & 2.32 \\
5156 & 20.5 & 3.56 & 10.5 & 2.44 \\
5172 & 25.5 & 4.13 & 11.5 & 2.55 \\
5255 & 20 & 3.51 & 10 & 2.38 \\
5257 & 21 & 3.62 & 11 & 2.49 \\
5271 & 25 & 4.07 & 11 & 2.49 \\
5273 & 26 & 4.18 & 12 & 2.61 \\
5275 & 27 & 4.29 & 13 & 2.72 \\
5353 & 19 & 3.39 & 9 & 2.27 \\
5355 & 20 & 3.51 & 10 & 2.38 \\
5371 & 25 & 4.07 & 11 & 2.49 \\
5373 & 26 & 4.18 & 12 & 2.61 \\
5375 & 27 & 4.29 & 13 & 2.72 \\
\bottomrule
\end{tabular}
\end{table}

\subsection{Expert Evaluation Coefficient} 

Observing Table \uppercase\expandafter{\romannumeral6}, this paper finds that when the number of somersault and twisting is the same, even if the take-off mode is different, the basic difficulty coefficient of diving action corresponding to different codes is still the same. (such as 105B, 205B, 305B, 405B four actions, air movements are 2.5 somersaults in pike position, the take-off method is different, but the basic difficulty coefficient is the same). Although the modeling and analysis of each diving action is relatively reasonable and accurate, in order to make the new diving difficulty coefficient rule more realistic, this paper intends to introduce the expert evaluation coefficient \emph{EA} to correct the basic difficulty coefficient with the help of expert experience. After a long period of observation of the diving competition, the expert could clearly understand the influence of different take-off modes on the diving difficulty coefficient. This paper proposes to combine the expert's observation experience with the old rule formulation experience to formulate the expert evaluation coefficient to ensure that the difficulty coefficient law of different take-off modes under the new difficulty coefficient rule is close to the old rule. Taking the X05 series action as an example, the basic difficulty coefficients of 105B, 205B, 305B, and 405B before correction are identical. After the revision of the expert evaluation coefficient based on the old rule experience, a new action difficulty coefficient \emph{DD'}(\emph{DD'=DD''×EA}) is obtained, as shown in Table \uppercase\expandafter{\romannumeral7}. The action difficulty coefficient corrected by the expert evaluation coefficient could fully reflect the influence of the take-off mode on the action difficulty coefficient, and the law is the same as the original action difficulty coefficient. 

\begin{table*}[!htbp]
\centering
\caption{Expert evaluation coefficient of X05 series action}
\label{T4-5}
\begin{tabular}{c|cccc|cccc}
\toprule
 &\multicolumn{4}{c}{B}&\multicolumn{4}{|c}{C} \\
\midrule
Action code & \emph{DD''} & \emph{EA} &\tabincell{c}{Original\\difficulty\\coefficient}& \emph{DD'} & \emph{DD''} & \emph{EA} &\tabincell{c}{Original\\difficulty\\coefficient}& \emph{DD'} \\
\midrule
105 & 3.23 & 0.78 & 2.3 & 2.53 & 2.1 & 1 & 2.1 & 2.1 \\
205 & 3.23 & 0.97 & 2.9 & 3.13 & 2.1 & 1.29 & 2.7 & 2.7 \\
305 & 3.23 & 1 & 3.0 & 3.23 & 2.1 & 1.33 & 2.8 & 2.8 \\
405 & 3.23 & 0.94 & 2.8 & 3.03 & 2.1 & 1.19 & 2.5 & 2.5 \\
\bottomrule
\end{tabular}
\end{table*}

Although there are certain subjective factors in the expert evaluation coefficient, in this way, experts can be hired to correct and re-engineer the difficulty coefficient according to the use in the practice process, so that it is more suitable for the actual game needs and better serve the game.

\subsection{New Diving Action Difficulty Coefficient Rules} 

Through the above work, this paper determines the basic difficulty coefficient \emph{DD''} according to the completion time of each action corresponding to different action codes. The difficulty of getting the same score in the same diving action for athletes of different body size is different, and it is necessary to introduce the body shape correction coefficient \emph{BC} to correct this difference, so that the corrected action difficulty coefficient truly reflects the difficulty of the specific player to complete a specific action. So this paper specifies the new action difficulty coefficient \emph{DD=DD''×EA×BC}. The new difficulty coefficient \emph{DD} developed in this paper is shown in Table \uppercase\expandafter{\romannumeral8}, where \emph{BC} could be obtained by looking up Table \uppercase\expandafter{\romannumeral4} and Table \uppercase\expandafter{\romannumeral5} according to the athlete's body shape correction index \emph{Mh$^{2}$}. 

\begin{table*}[!htbp]
\centering
\caption{The 10-meter platform action coefficient}
\label{T4-6}
\begin{tabular}{c|cccc|cccc}
\toprule
 &\multicolumn{4}{c}{B}&\multicolumn{4}{|c}{C} \\
\midrule
Action code & \emph{DD''} & \emph{EA} &\tabincell{c}{Original\\difficulty\\coefficient}& \emph{DD} & \emph{DD''} & \emph{EA} &\tabincell{c}{Original\\difficulty\\coefficient}& \emph{DD} \\
\midrule
105 & 3.23 & 0.78 & 2.3 & 2.53\emph{BC} & 2.1 & 1 & 2.1 & 2.1\emph{BC} \\
107 & 4.01 & 0.83 & 3.0 & 3.31\emph{BC} & 2.44 & 1 & 2.7 & 2.44\emph{BC} \\
109 & 4.8 & 0.85 & 4.1 & 4.1\emph{BC} & 2.78 & 1 & 3.3 & 2.78\emph{BC} \\
1011 & 5.59 & -- & -- & -- & 3.11 & 1 & 4.7 & 3.11\emph{BC} \\
205 & 3.23 & 0.97 & 2.9 & 3.13\emph{BC} & 2.1 & 1.29 & 2.7 & 2.7\emph{BC} \\
207 & 4.01 & 0.98 & 3.6 & 3.91\emph{BC} & 2.44 & 1.25 & 3.3 & 3.04\emph{BC} \\
209 & 4.8 & 0.94 & 4.5 & 4.5\emph{BC} & 2.78 & 1.18 & 4.2 & 3.28\emph{BC} \\
305 & 3.23 & 1 & 3.0 & 3.23\emph{BC} & 2.1 & 1.33 & 2.8 & 2.8\emph{BC} \\
307 & 4.01 & 1 & 3.7 & 4.01\emph{BC} & 2.44 & 1.29 & 3.4 & 3.14\emph{BC} \\
309 & 4.8 & 1 & 4.8 & 4.8\emph{BC} & 2.78 & 1.29 & 4.5 & 3.58\emph{BC} \\
405 & 3.23 & 0.94 & 2.8 & 3.03\emph{BC} & 2.1 & 1.19 & 2.5 & 2.5\emph{BC} \\
407 & 4.01 & 0.95 & 3.5 & 3.81\emph{BC} & 2.44 & 1.20 & 3.2 & 2.94\emph{BC} \\
409 & 4.8 & 0.92 & 4.4 & 4.4\emph{BC} & 2.78 & 1.14 & 4.1 & 3.18\emph{BC} \\
5154 & 3.45 & 0.82 & 3.3 & 2.82\emph{BC} & 2.32 & 1 & 3.1 & 2.32\emph{BC} \\
5156 & 3.56 & 0.93 & 3.8 & 3.32\emph{BC} & 2.44 & 1.16 & 3.6 & 2.82\emph{BC} \\
5172 & 4.13 & 0.96 & 3.6 & 3.98\emph{BC} & 2.55 & 1 & 3.3 & 2.55\emph{BC} \\
5255 & 3.51 & 0.89 & 3.6 & 3.12\emph{BC} & 2.38 & 1.1 & 3.4 & 2.62\emph{BC} \\
5257 & 3.62 & 1 & 4.1 & 3.62\emph{BC} & 2.49 & 1.25 & 3.9 & 3.12\emph{BC} \\
5271 & 4.07 & 0.98 & 3.2 & 3.97\emph{BC} & 2.49 & 1 & 2.9 & 2.49\emph{BC} \\
5273 & 4.18 & 1 & 3.8 & 4.18\emph{BC} & 2.61 & 1.05 & 3.5 & 2.75\emph{BC} \\
5275 & 4.29 & 1 & 4.2 & 4.29\emph{BC} & 2.72 & 1.18 & 3.9 & 3.2\emph{BC} \\
5353 & 3.39 & 0.83 & 3.3 & 2.82\emph{BC} & 2.27 & 1.02 & 3.1 & 2.32\emph{BC} \\
5355 & 3.51 & 0.92 & 3.7 & 3.22\emph{BC} & 2.38 & 1.14 & 3.5 & 2.72\emph{BC} \\
5371 & 4.07 & 1 & 3.3 & 4.07\emph{BC} & 2.49 & 1.04 & 3.0 & 2.59\emph{BC} \\
5373 & 4.18 & -- & -- & -- & 2.61 & 1.25 & 3.6 & 3.25\emph{BC} \\
5375 & 4.29 & -- & -- & -- & 2.72 & 1.34 & 4.0 & 3.65\emph{BC} \\
\bottomrule
\end{tabular}
\end{table*}

Through the comparison of the old and new difficulty coefficients, it could be found that the new action difficulty coefficient table is consistent with the old action difficulty coefficient table in some rules. These rules are:

1. When the take-off mode, the number of somersault and twisting are the same, the difficulty coefficient of the somersault in pike position of the action code \emph{B} is higher than the somersault in tuck position of the action code \emph{C}.

2. When the somersault mode, the number of somersault and twisting are the same, and the athlete's take-off mode and the direction of the somersault are different, the difficulty of the action is different. The difficulty arrangement of the first digit of the action code from high to low is 3,2,4,1.

3. The more the number of somersault and twisting, the higher the difficulty coefficient of the corresponding action.

It can be seen that the rules of the old and new action difficulty coefficients are consistent in many aspects. These consistent laws are in line with the analysis of this paper and people's perceptions. Still, the rules of the new action difficulty coefficient and the old action difficulty coefficient also differ in some respects, including:

1. The new difficulty coefficient rule introduces the body shape correction coefficient \emph{BC}. Considering the influence of the body shape correction index \emph{Mh$^{2}$} on the action difficulty, the athlete's body shape will affect the time required to complete an action, and thus affect the difficulty coefficient of a set of actions. Therefore, this paper introduces the body shape correction coefficient \emph{BC} on the basis of the old difficulty coefficient rule to regulate the influence of the body size on the action difficulty.

2. In the new difficulty coefficient rule, the difference in the difficulty coefficient between the somersault in pike position and the somersault in tuck position is increased. Assume that the athlete's body shape correction index \emph{Mh$^{2}$} is close to the average level, and the body shape correction coefficient \emph{BC=1}. For this athlete, in the old action difficulty coefficient rule, the difficulty difference between 105B and 105C is 0.2, and in the new action difficulty coefficient rule, the difficulty difference between 105B and 105C is 0.43, the difficulty difference is increased, and this rule applies to most action codes. The increase in the difficulty difference is due to the change in the quantitative method of the difference between the difficulty of these two actions. In the old action difficulty coefficient rule, the difficulty difference between these two actions is the fixed value, and the maximum difference is 0.3; in the new difficulty coefficient rule, the difficulty difference is mainly reflected in the time ratio of these two actions. According to the model of this paper, the ratio of the completion time of these two actions is \emph{7:3}, that is, the difficulty of somersault in pike position is 7/3 times the difficulty of somersault in tuck position. The change of the quantitative method of the difficulty difference between these two actions has led to an increase in the difficulty coefficient difference of the two actions in the new difficulty coefficient rule.

\section{Simulation} 

This paper takes the actions of the female diving athletes X and Y in the 2017 FINA World Championships as an example to illustrate the use of the new diving difficulty coefficient rule and the influence of the change of rules on the athlete's diving score.

X completed four actions of 107B, 407C, 207C, and 5353B in the game. Taking the 107B action as an example, look up Table \uppercase\expandafter{\romannumeral8} and get the new difficulty coefficient of the 107B action is 3.31\emph{BC}. The height of X is 1.62m and the weight is 45kg. Therefore, the body shape correction index \emph{Mh$^{2}$} is \emph{$118.098kg\cdot m^{2}$}. Looking up Table \uppercase\expandafter{\romannumeral4}, the body shape correction index \emph{BC} of X is 0.93, so the 107B action difficulty coefficient of X is \emph{3.31×0.93=3.08}. In the game, X's 107B action score is 26 points, so the total score of X's 107B action is \emph{26×3.08=80.08} points. Repeat the above process to get the scores of each action of X, as shown in Table \uppercase\expandafter{\romannumeral9}. 

\begin{table*}[htbp]
\centering
\caption{Scores of each action of X}
\label{T5-1}
\begin{tabular}{ccccccc}
\toprule
Action number & Action code & Action score & Old difficulty coefficient &\tabincell{c}{Total score\\under old rule}& New difficulty coefficient &\tabincell{c}{Total score\\under new rule} \\
\midrule
1 & 107B & 26 & 3.0 & 78 & 3.08 & 80.08\\
2 & 407C & 28.5& 3.2 & 91.2 & 2.73 & 77.81\\
3 & 207C & 19.5 & 3.3 & 64.35 & 2.83 & 55.19\\
4 & 5353B & 24 & 3.3 & 79.2 & 2.62 & 62.88\\
\bottomrule
\end{tabular}
\end{table*}

It is calculated that the total score of X’s four actions under the old rules is 312.75 points, and the total score under the new rules is 275.96 points. Compared to the old rules, X has a lower total score under the new rules. The reason is that the difficulty coefficients of 407C, 207C, and 5253B under the new rule are reduced. In the old rules, the difficulty coefficient of 407C and 207C is higher than 107B. According to the analysis of this article, the ratio of completion time of full somersault in pike position and full somersault in tuck position is \emph{7:3}. When the number of somersault is the same, the difficulty of somersault in pike position is higher than that of somersault in tuck position, so the difficulty of the 407C and 207C under the new rules is lower than the 107B, which is also 3.5 somersaults, the difficulty coefficient of these two actions is much lower than the old one. Under the old rules, the difficulty coefficient of the 5353B action is the highest, including 2.5 somersaults and 1.5 twisting. According to the analysis of this article, the ratio of completion time of full somersault in pike position and full twisting is \emph{7:1}, the difficulty of somersault in pike position is much higher than twisting action, so the difficulty of a set of twisting somersault action is mainly reflected in the number of somersault, Although the 5353B has complicated actions, the number of somersault is small, so the difficulty coefficient of 5353B under the new rules is much lower than that of the old rules.

Y also completed the four actions of 107B, 407C, 207C, 5353B in the game. Taking the 107B action as an example, look up Table \uppercase\expandafter{\romannumeral8} and get the new difficulty coefficient of the 107B action is 3.31\emph{BC}. The height of Y is 1.64m and the weight is 50kg. Therefore, the body shape correction index \emph{Mh$^{2}$} is \emph{$134.48kg\cdot m^{2}$}. Looking up Table \uppercase\expandafter{\romannumeral4}, the body shape correction index \emph{BC} of Y is 1.22, so the 107B action difficulty coefficient of Y is \emph{3.31×1.22=4.04}. In the game, the 107B action score of Y is 24.5 points, so the total score of 107B action of Y is \emph{24.5×4.04=98.98} points. Repeat the above process to get the scores of each action of Y, as shown in Table \uppercase\expandafter{\romannumeral10}. 

\begin{table*}[htbp]
\centering
\caption{Scores of each action of Y}
\label{T5-2}
\begin{tabular}{ccccccc}
\toprule
Action number & Action code & Action score & Old difficulty coefficient &\tabincell{c}{Total score\\under old rule}& New difficulty coefficient &\tabincell{c}{Total score\\under new rule} \\
\midrule
1 & 107B & 24.5 & 3.0 & 73.5 & 4.04 & 98.98\\
2 & 407C & 25.5& 3.2 & 81.6 & 3.59 & 91.55\\
3 & 207C & 25 & 3.3 & 82.5 & 3.71 & 92.75\\
4 & 5353B & 25.5 & 3.3 & 84.15 & 3.44 & 87.72\\
\bottomrule
\end{tabular}
\end{table*}

It is calculated that the total score of the four actions of Y under the old rule is 321.75, and the total score under the new rule is 371. Under the old rules, the total scores of the four actions of X and Y are similar. But under the new rules, Y's total score is nearly 100 points higher than X. The reason is that the body shape correction index \emph{Mh$^{2}$} of Y is higher than X, and Y has a body shape disadvantage compared with X. Therefore, the body shape correction coefficient \emph{BC} of Y is higher than X, and the total score of Y after correction is much higher than X, which fully reflects the effect of body shape correction coefficient to reduce the body size advantage of the skinny athletes.

It can be seen from the above analysis that the difficulty coefficient of somersault in tuck position and twisting somersault under the new rule is greatly reduced compared with the old rule, and selecting these actions in the game will lose the advantage of the difficulty coefficient. In addition, the body shape correction coefficient plays a role in reducing the body size advantage of the skinny athletes, making the rules of the game more fair.

\section{Conclusion} 

In this paper, the multi-rigid-body model of the human body is established based on the problem of the body shape correction coefficient of the 10-meter platform diving. On this basis, the relationship between the time for the athletes to complete each diving action and the height \emph{h} and weight \emph{M} of the athlete is obtained. And the body shape correction index is determined to be \emph{Mh$^{2}$}, which confirms the necessity of setting the body shape correction coefficient. Then, the function relationship between the body shape correction coefficient \emph{BC} and the body shape correction index \emph{Mh$^{2}$} is established by the Lagrange Interpolation Polynomial, and the body shape correction coefficient scheme is formulated accordingly. Finally, using the model established and the conclusions obtained in the previous paper, a new rule of action difficulty coefficient is formulated, and the expert evaluation coefficient \emph{EA} is introduced to correct the new rule so that the new action difficulty coefficient rule is more realistic. The model proposed in this paper is simple and easy to understand, and maintains the appropriate model complexity. The model is in line with the reality and provides a simple and reliable new idea for the formulation of diving difficulty coefficient. In addition, the new action difficulty coefficient rule introduces the expert evaluation coefficient, which makes the model variability and flexibility in the future development process. The action difficulty coefficient could be corrected according to the change of the real situation in real time, so that the rule could better serve the game and ensure the fairness of diving.

However, the research work in this paper is quite rough in some aspects: the diving process assumed in this paper does not consider the tilting action after the take-off, which is slightly different from the actual diving process and a certain deviation between the relationship of the completion time of each diving action and the height and weight of the athlete and the actual situation; when the Lagrange Interpolation Polynomial is used to solve the functional relationship between the body shape correction coefficient and the body shape correction index, the number of interpolation points is small due to the limitation of the data source, and there are certain subjective factors in the formulation of interpolation point coordinates, resulting in a low degree of functional relationship between the body shape correction coefficient and the body shape correction index and a certain error; the basic action difficulty coefficient is only determined according to the completion time of each diving action, and does not consider the impact of the athlete's take-off mode on the action completion time, and uses the linear model to solve the basic difficulty coefficient of each action, which may have some gaps with the reasonable difficulty coefficient of each action. Even if it can be corrected by the expert evaluation coefficient, it is necessary to make a long-term modification of the expert evaluation coefficient in order to make the new action difficulty coefficient truly fit. These aspects require more detailed research in the future work, so that the new diving difficulty coefficient rules are more reasonable and perfect.

\section*{Acknowledgment}

I would like to thank School of mechanical engineering, Shanghai Jiao Tong University.

%
\begin{IEEEbiography}[{\includegraphics[width=1in,height=1.25in,clip,keepaspectratio]{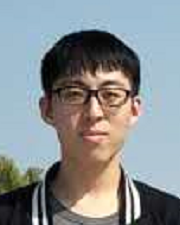}}]{Yan-Xin Sun} received the B.S. degree in Mechanical Engineering from Harbin Institute of Technology in 2017, and is currently a Master candidate with the School of Mechanical Engineering and Institute of Manufacturing and Equipment Automation, Shanghai Jiao Tong University, Shanghai 200240,China, from Fall 2017. 

His current research interests include electric discharge machining, CNC system, Computer integrated manufacturing System, mechanical Simulation and Modeling.
\end{IEEEbiography}

\begin{IEEEbiography}[{\includegraphics[width=1in,height=1.25in,clip,keepaspectratio]{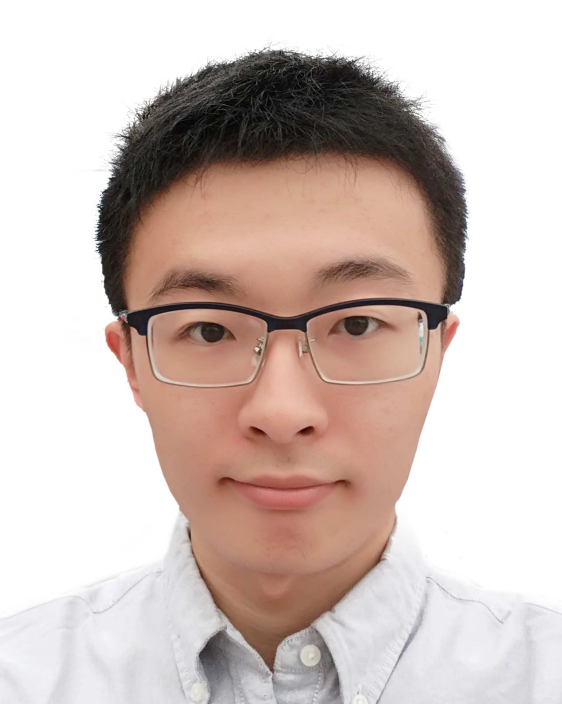}}]{Zhao-Hui Sun}(M'17) is currently a Master candidate with the School of Mechanical Engineering and Institute of Intelligent Manufacturing, Shanghai Jiao Tong University, Shanghai 200240,China, from Fall 2017. 

His current research interests include computational intelligence, data mining, industrial artificial intelligence, industrial blockchain, industrial big data, and their applications in smart manufacturing and Enterprise Informatics.

\end{IEEEbiography}

\end{document}